# Enhanced Visibility of $MoS_2$, $MoSe_2$, $WSe_2$ and Black-Phosphorus: Making Optical Identification of 2D Semiconductors Easier

Gabino Rubio-Bollinger [1,2], Ruben Guerrero [3], David Perez de Lara [3], Jorge Quereda [1], Luis Vaquero-Garzon [3], Nicolas Agraït [1,2,3], Rudolf Bratschitsch [4] and
Andres Castellanos-Gomez [3,*]

[1] Dpto. de Física de la Materia Condensada, Universidad Autónoma de Madrid, 28049 Madrid, Spain.

[2] Condensed Matter Physics Center (IFIMAC), Universidad Autónoma de Madrid,
E-28049 Madrid, Spain.

[3] Instituto Madrileño de Estudios Avanzados en Nanociencia (IMDEA-nanociencia),
Campus de Cantoblanco, E-18049 Madrid, Spain.

[4] Institute of Physics and Center for Nanotechnology, University of Münster, D-48149 Münster, Germany.
**\*** Author to whom correspondence should be addressed; E-Mail: andres.castellanos@imdea.org

**Abstract:** We explore the use of $Si_3N_4$/Si substrates as a substitute of the standard $SiO_2$/Si substrates employed nowadays to fabricate nanodevices based on 2D materials. We systematically study the visibility of several 2D semiconducting materials that are attracting a great deal of interest in nanoelectronics and optoelectronics: $MoS_2$, $MoSe_2$, $WSe_2$ and black-phosphorus. We find that the use of $Si_3N_4$/Si substrates provides an increase of the optical contrast up to a 50%–100% and also the maximum contrast shifts towards wavelength values optimal for human eye detection, making optical identification of 2D semiconductors easier.

**1. Introduction**

Mechanical exfoliation has proven to be a very powerful tool to isolate two-dimensional material out of bulk layered crystal [1]. The produced atomically thin layers, however, are randomly deposited on the substrate surface and are typically surrounded by thick crystal which hampers the identification of the thinner material. Optical microscopy is a perfect complement to mechanical exfoliation as is a reliable and non-destructive method and it allows distinguishing the atomically thick layers from their bulk-like counterparts. This technique is valid despite the reduced thickness of 2D materials. They can be seen through an optical microscope with the naked eye, because of the wavelength dependent reflectivity of the dielectric/2D material system [2–6]. This dependence can be exploited to easily identify and isolate 2D material single layer flakes by modifying the substrate dielectric thickness and permittivity. In addition to increasing the visibility, the use of different substrate materials may improve the performance of the produced devices if the chosen substrate has good dielectric properties.

In this work we systematically study the visibility of several 2D materials with potential applications in electronics and optoelectronics, such as $MoS_2$, $MoSe_2$, $WSe_2$ and black-phosphorus (BP) [7–11]. The performed experiments and analysis are general, and can be applied to any kind of 2D materials. Here, we explore the use of silicon nitride ($Si_3N_4$), a high $k$ dielectric material ($\kappa \sim 7$) commonly used in the semiconductor industry, as a substitute of the silicon oxide layer, which is almost exclusively used nowadays to fabricate nanodevices based on 2D materials. We show how the use of silicon nitride strongly enhances the optical contrast of 2D semiconductors, making the identification of ultrathin sheets easier. Moreover, by using a $Si_3N_4$ spacer layer of 75 nm in thickness, the optical contrast reaches its maximum value at a wavelength of 550 nm (which is the optimal wavelength detection of the human eye) [12], while 285 nm of $SiO_2$ spacer layer (the standard in graphene and $MoS_2$ research nowadays) has its maximum contrast value at 460 nm, in the deep-blue/violet part of the visible spectrum.

Apart from the enhanced visibility, the use of $Si_3N_4$ as spacer layer also has the potential to improve the electrical performance of nanoelectronic devices due to its high dielectric constant (almost twice that of $SiO_2$) that can help to screen Coulomb scatterers and, thus, to improve the mobility [13]. Additionally, the use of $Si_3N_4$ do not present any disadvantage with respect to $SiO_2$ layers in terms of processing as $Si_3N_4$ is compatible with most common semiconductor industry processes. Moreover, $Si_3N_4$ substrates can be used for other fabrication processes different than mechanical exfoliation, such as CVD [14–16] (because of its high thermal stability) and inkjet printing [17–19] as its surface chemistry has the potential to be tuned using similar recipes to those used for $SiO_2$ substrates. In summary due to the good dielectric performance of $Si_3N_4$ and its deposition compatibility with other semiconductor industry processes, we believe that the use of $Si_3N_4$ as spacer layer for 2D semiconductor applications will become popular in the near future.

## 2. Experimental Section

Two-dimensional semiconductor samples are prepared using a recently developed deterministic transfer process [20]. First, we mechanically exfoliate bulk $MoS_2$, $MoSe_2$, $WSe_2$ or black phosphorous using clear Nitto tape (SPV 224). Bulk crystals were synthetic (grown by vapor transport method) in all cases except the $MoS_2$ crystal that was obtained from naturally occurring molybdenite (Moly Hill mine, Quebec, QC, Canada). The freshly cleaved flakes are then deposited onto a viscoelastic poly-dimethylsiloxane (PDMS) substrate. Subsequently, the flakes are transferred onto two different silicon substrates: one with a 285 nm thick $SiO_2$ oxide layer on top and another one with a 75 nm thick $Si_3N_4$ layer. The latter thickness was chosen after the theoretical analysis explained in Section 3 in order to maximize the contrast at a wavelength of 550 nm.

Few-layer flakes are identified under an optical microscope (Nikon Eclipse LV100) and the number of layers is determined by a combination of quantitative optical microscopy and contact mode atomic force microscopy (used instead of tapping mode to avoid artifacts in the thickness determination). The optical properties of the nanosheets have been studied with a modification of a home-built hyperspectral imaging setup, described in detail in Reference [21].

## 3. Results and Discussion

### 3.1. Optical Contrast Calculation

In order to evaluate the potential of $Si_3N_4$ to enhance the optical visibility of 2D semiconductors we have first calculated the optical contrast of monolayer $MoS_2$, $MoSe_2$ and $WSe_2$ as function of the illumination wavelength for substrates with $Si_3N_4$ and $SiO_2$ layers of different thickness. The model is based on the Fresnel law and more details can be found in the literature [2,3,22–25]. Briefly, the optical contrast of atomically thin materials is due to a combination of: (1) interference between the reflection paths that originate from the interfaces between the different media and (2) thickness dependent transparency of the 2D material that strongly modulates the relative amplitude of the different reflection paths. These two effects combined lead to color shifts (dependent on the thickness of the 2D material) that can be appreciated by eye. Figure 1 displays colormaps that represent the optical contrast (defined as $C = (I_{flake} - I_{substrate})/(I_{flake} + I_{substrate})$) as a function of the illumination wavelength (vertical axis) and the thickness of the dielectric layer (horizontal axis). The references employed to extract the refractive indexes for the different materials employed in the calculation of the optical contrast are summarized in Table 1. One can clearly see how the optical contrast for $Si_3N_4$ substrates is much higher than for $SiO_2$. Moreover, substrates with a 75 nm $Si_3N_4$ layer have a maximum contrast at a wavelength around 550 nm, which is optimal for human eye detection. The strong optical contrast enhancement observed for 75 nm thick $Si_3N_4$ layers on Si substrates can be easily understood as the combination of thickness and refractive index of this layer makes it an almost perfect anti-reflective coating. An optimal anti-reflective coating should have a refractive index $n_{ar} = \sqrt{n_{air} \cdot n_{Si}} \approx 2$ (very similar to the $n_{Si3N4}$ in the visible part of the spectrum) and a thickness $d_{ar} = \lambda/4n_{ar} \approx 65 - 75 nm$ (to minimize the reflection within the visible part of the spectrum). In fact, the reflectivity of a 75 nm thick $Si_3N_4$ layer on Si is almost zero in the visible range of spectrum. Therefore, the contrast of the 2D materials is enhanced, because their surrounding substrate almost does not reflect any light.

As one can see from Figure 1, the optical contrast on $Si_3N_4/Si$ substrates is more sensitive to small variations of the spacing layer than the $SiO_2/Si$ substrates. While for $SiO_2/Si$ substrates can present the $SiO_2$ thickness variations of ~ ±10 nm, the $Si_3N_4$ layer thickness should present variations within ~±3 nm to avoid substantial contrast variations within the sample. We address the readers to the Supporting Information to see a comparison between two horizontal linecuts along 550 nm wavelength in panel (a) and (b).

The result of the calculations displayed in Figure 1 illustrates the potential of $Si_3N_4$ spacer layers with a thickness of 50 nm–100 nm to enhance the optical contrast significantly with respect to conventionally used $SiO_2$ spacer layers. Therefore, we have explored the potential of $Si_3N_4$ in 2D semiconductor research by experimentally studying the optical contrast of several 2D semiconductors ($MoS_2$, $MoSe_2$, $WSe_2$ and black phosphorus) on silicon substrates with a 75 nm thick $Si_3N_4$ layer (IDB Technologies Ltd, Whitley, Wiltshire, UK). We also fabricated samples on $Si/SiO_2$ substrates with 285 nm $SiO_2$ in order to compare the measured optical contrast with the most extended dielectric layer used in 2D materials research nowadays.

## 3.2. Hyperspectral Imaging

The optical contrast is measured at different illumination wavelengths with a modified hyperspectral imaging setup described in Reference [21]. The sample is illuminated with monochromatic light with the help of a monochromator. The measurement is carried out by sweeping the illumination wavelength from 450 nm to 900 nm in 5 nm steps, and acquiring an image with a monochrome camera at each wavelength step. The thickness of the studied flakes has been determined by atomic force microscopy prior to the hyperspectral imaging measurements (see Figure 2). Raman spectroscopy or photoluminescence can be also used to characterize and to determine the thickness of the exfoliated flakes on $Si_3N_4$ surfaces [30], see Supporting Information for Raman spectra acquired for $MoS_2$ flakes on a 75 nm $Si_3N_4$/Si substrate and a comparison with the spectra reported for flakes on 285 nm $SiO_2$/Si substrates [31,32].

Figure 3 shows the obtained optical contrast maps of $MoS_2$ flakes with a single-layer region (highlighted in the Figure with "1 L") on a 285 nm $SiO_2$/Si substrate (a) and on a 75 nm $Si_3N_4$/Si substrate (b) under illumination with different wavelengths: 500 nm, 550 nm, 600 nm, 650 nm, 700 nm and 750 nm. The comparison between the results obtained for the $SiO_2$ and $Si_3N_4$ layers illustrates how the optical contrast of monolayer $MoS_2$ on $SiO_2$ is weaker within the visible part of the spectrum, whereas for $Si_3N_4$ around 500–600 nm the monolayer contrast reaches the highest value.

## 3.3. Wavelength Dependent Optical Contrast

From the contrast maps at different wavelengths one can extract the wavelength dependence of the optical contrast for flakes with different thicknesses. Figure 4 summarizes the contrast *vs.* wavelength dependence measured for $MoS_2$, $MoSe_2$, $WSe_2$ and black phosphorus on both substrates. For all the studied materials the optical contrast is enhanced on substrates with $Si_3N_4$ by a 50%−100%. The wavelength with the maximum optical contrast is also shifted: while on $SiO_2$/Si substrates it is ~650 nm, on $Si_3N_4$ the maximum contrast is at ~550 nm.

## 4. Conclusions

In summary, we have explored the use of $Si_3N_4$ as dielectric layer for 2D semiconductor research. We systematically studied the optical contrast of several 2D semiconductors ($MoS_2$, $MoSe_2$, $WSe_2$ and black phosphorus) on silicon substrates with 75 nm of $Si_3N_4$ spacer layers, which according to our calculations should substantially enhance the optical contrast. We experimentally demonstrated the optical contrast enhancement due to 75 nm $Si_3N_4$/Si substrates by measuring the optical contrast in the range of 450 nm to 900 nm by hyperspectral imaging. We compared the measured contrast to that acquired for samples fabricated on the standard 285 nm $SiO_2$/Si substrates, finding an increase of the optical contrast up to a 50%–100%. The maximum contrast also shifts in wavelength towards wavelength values optimal for human eye detection. The obtained results provide a way of improving optical identification of single layers of 2D materials.

## Acknowledgments

AC-G acknowledges financial support from the BBVA Foundation through the fellowship "I Convocatoria de Ayudas Fundacion BBVA a Investigadores, Innovadores y Creadores Culturales", from the MINECO (Ramón y Cajal 2014 program, RYC-2014-01406) and from the MICINN (MAT2014-58399-JIN). R.G. acknowledges financial support from the AMAROUT-Marie Curie program. We also acknowledge funding from the projects MAT2014-57915-R (MINECO) and MAD2D project S2013/MIT-3007 (Comunidad Autónoma de Madrid).

**Table 1.** Summary of the references with the refractive index values necessary for the calculation of the optical contrast.

| Material | Reference |
|---|---|
| $SiO_2$ | [26] |
| $Si_3N_4$ | [27] |
| $MoS_2$ | [28] |
| $MoSe_2$ | [28] |
| $WSe_2$ | [29] |

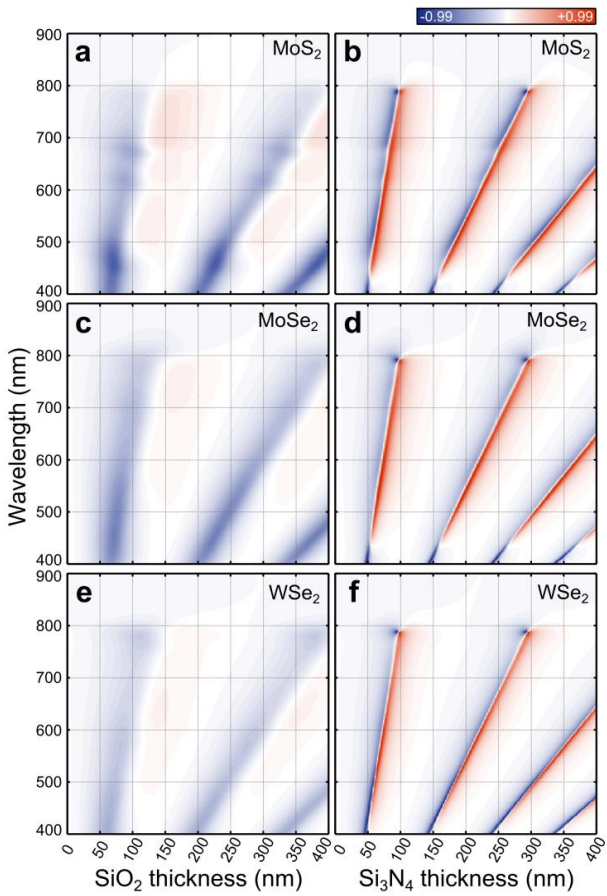

**Figure 1.** Calculated optical contrast as a function of the illumination wavelength and spacer layer thickness for monolayer $MoS_2$ (**a,b**), $MoSe_2$ (**c,d**) and $WSe_2$ (**e,f**) for substrates with $SiO_2$ (left panel) and $Si_3N_4$ (right panel) spacer layers.

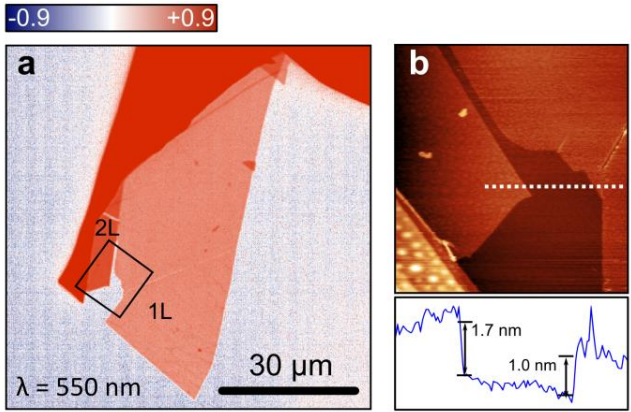

**Figure 2.** (**a**) Contrast map of a MoS$_2$ flake deposited onto a 75 nm Si$_3$N$_4$/Si substrate under illumination with 550 nm wavelength; (**b**) Topographic atomic force microscopy image acquired on the region highlighted with the square in (a), a topographic line profile is also shown below.

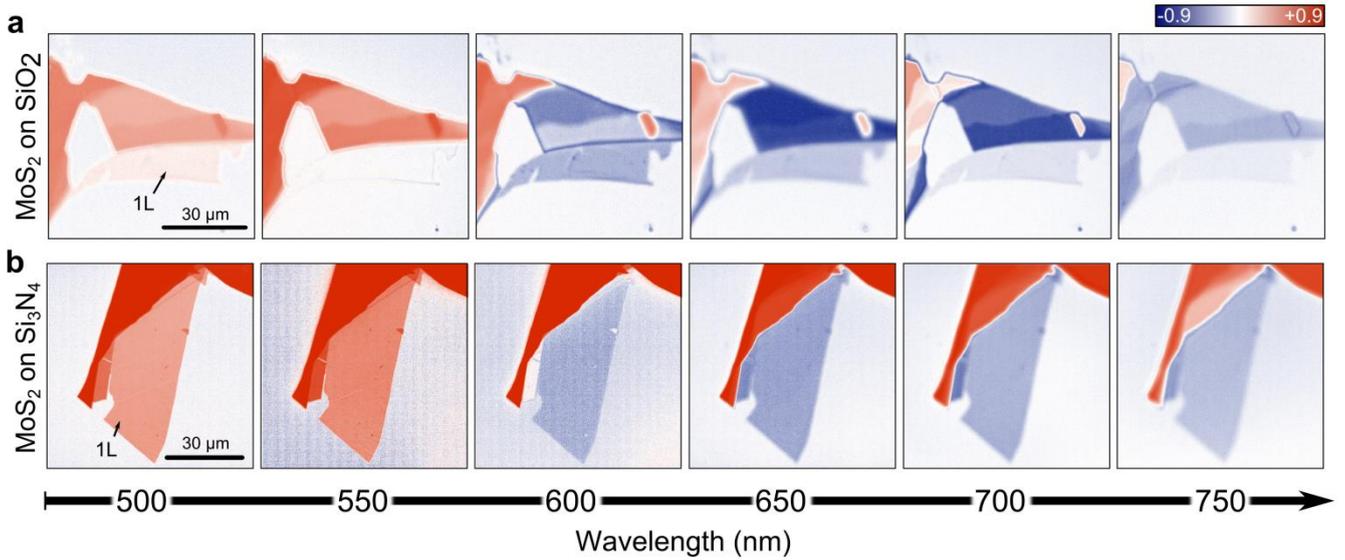

**Figure 3.** Contrast maps obtained by hyperspectral imaging of MoS$_2$ flakes on Si substrate with spacer layers of, (**a**) 285 nm of SiO$_2$ or (**b**) 75 nm of Si$_3$N$_4$, at different illuminating wavelengths. The contrast in the single layer case is maximum at 550 nm for Si$_3$N$_4$, whereas in the SiO$_2$ case is at 600 nm. It allows for the direct identification of single layer flakes.

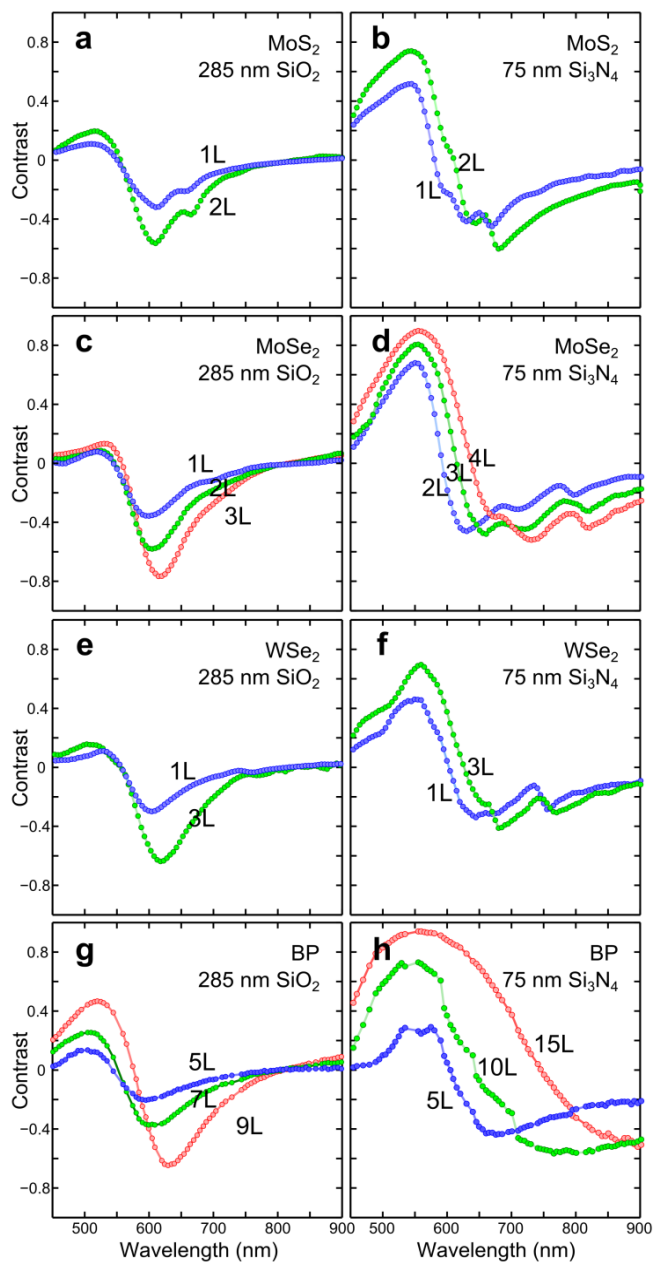

**Figure 4.** Wavelength dependence of the optical contrast measured for $MoS_2$, $MoSe_2$, $WSe_2$ and black phosphorus flakes of different thickness. (**a,b**) optical contrast of $MoS_2$ on 285 nm $SiO_2$/Si and 75 nm $Si_3N_4$/Si, respectively; (**c,d**) optical contrast of $MoSe_2$ on 285 nm $SiO_2$/Si and 75 nm $Si_3N_4$/Si, respectively; (**e,f**) optical contrast of $WSe_2$ on 285 nm $SiO_2$/Si and 75 nm $Si_3N_4$/Si, respectively; (**g,h**) optical contrast of black phosphorus on 285 nm $SiO_2$/Si and 75 nm $Si_3N_4$/Si, respectively.

# Supporting Information:

# Enhanced visibility of $MoS_2$, $MoSe_2$, $WSe_2$ and black-phosphorus: making optical identification of 2D semiconductors easier

**Gabino Rubio-Bollinger,[1,2] Ruben Guerrero,[3] David Perez de Lara,[3] Jorge Quereda,[1] Luis Vaquero-Garzon,[3] Nicolas Agraït,[1,2],3 Rudolf Bratschitsch[4] and Andres Castellanos-Gomez[3*]**

[1] Dpto. de Física de la Materia Condensada, Universidad Autónoma de Madrid, 28049 Madrid, Spain.

[2] Condensed Matter Physics Center (IFIMAC), Universidad Autónoma de Madrid, E-28049 Madrid, Spain.

[3] Instituto Madrileño de Estudios Avanzados en Nanociencia (IMDEA-nanociencia), Campus de Cantoblanco, E-18049 Madrid, Spain.

[4] Institute of Physics, University of Münster, D-48149 Münster (Germany).

\* Author to whom correspondence should be addressed; E-Mail: andres.castellanos@imdea.org ; Tel.: +34-91-299-8770

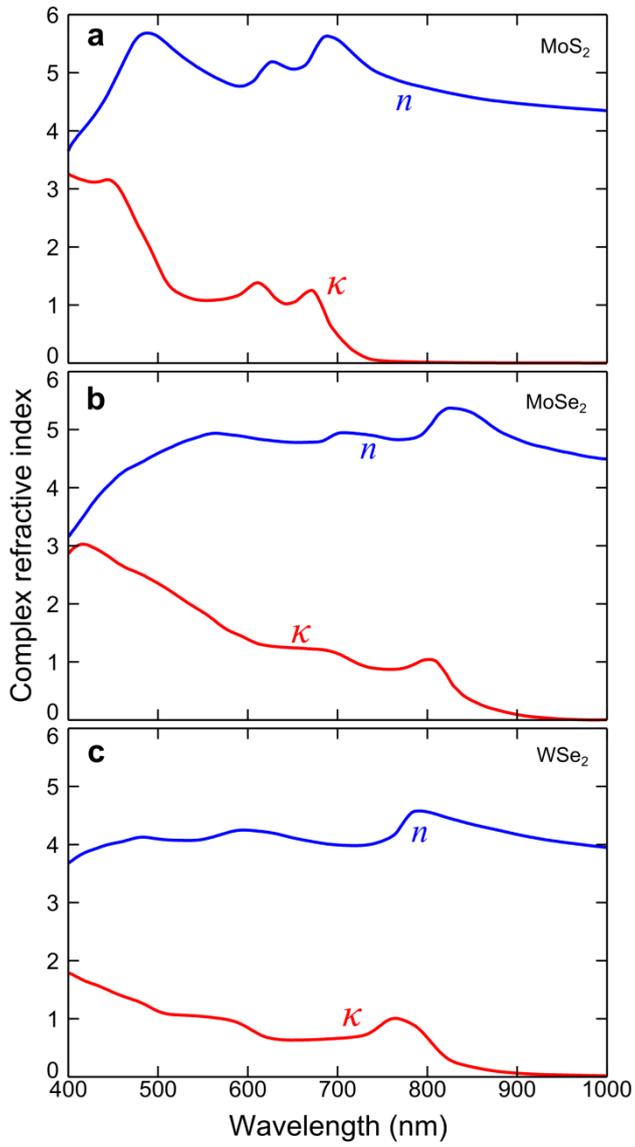

**Figure S1.** Real and imaginary parts of the refractive index of bulk $MoS_2$, $MoSe_2$ and $WSe_2$ extracted from the complex dielectric constants displayed in Ref. [22] and Ref. [23] of the main text. We display these values here to facilitate future calculations on these materials.

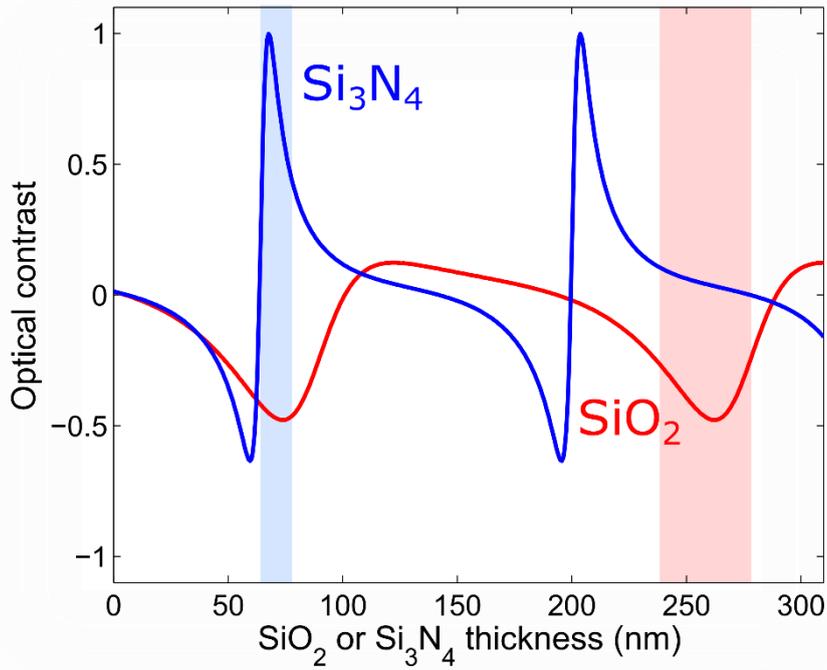

**Figure S2.** Optical contrast of a MoS$_2$ monolayer *vs.* thickness of the dielectric layer at a fixed illumination wavelength (550 nm). The regions highlighted with light color rectangles indicates the acceptable thickness variation of the Si$_3$N$_4$ (left) and SiO$_2$ (right) layers.

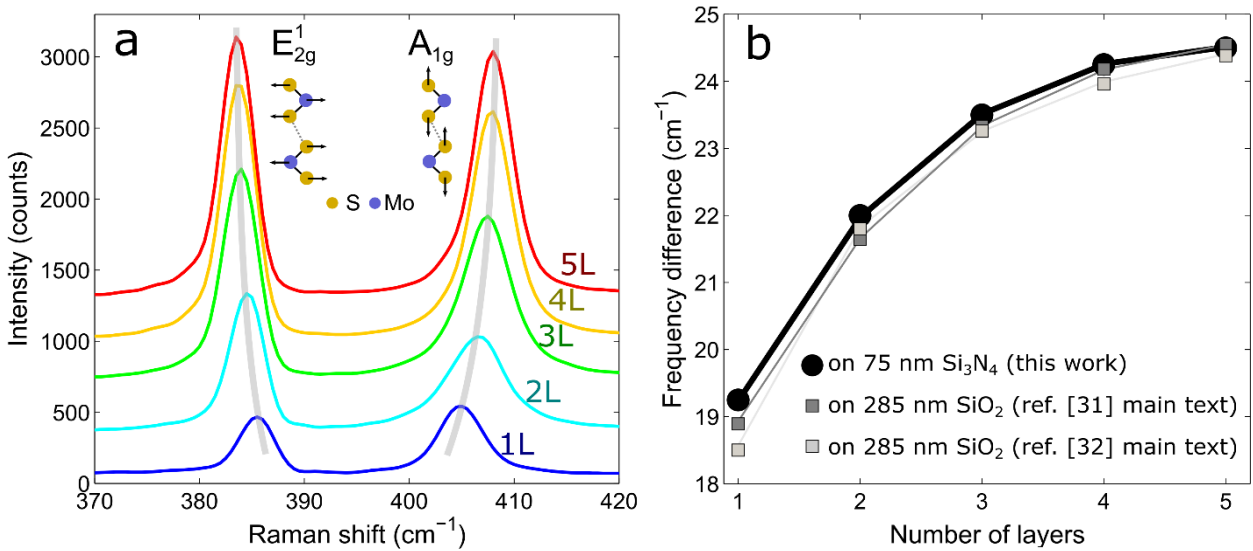

**Figure S3.** (a) Raman spectra measured on MoS$_2$ layers of different thickness transferred onto a 75 nm thick Si$_3$N$_4$/Si substrate. (b) The difference between the two Raman modes (E$^1_{2g}$ and A$_{1g}$) *vs.* the number of layers, extracted from (a). As a comparison the results reported for MoS$_2$ samples fabricated on 285 nm thick SiO$_2$/Si substrates have been also plotted (data extracted from Ref. [31] and Ref. [32] of the main text).